\documentclass[a4paper]{ISOStyle}
\usepackage{epsfig}

\begin{document}

\title{Photometric Calibrations for the SIRTF Infrared Spectrograph}

\author{P.\,W. Morris\inst{1,2} \and V. Charmandaris\inst{3} 
  \and T. Herter\inst{3} \and L. Armus\inst{1} \and J. Houck\inst{3} \and G. Sloan\inst{3}} 
  
  \institute{SIRTF Science Center, California Institute of Technology,
  1200 E. California Blvd., Pasadena, CA 91125
  \and NASA Herschel Science Center, California Institute of Technology,
  1200 E. California Blvd., Pasadena, CA 91125
  \and Department of Astronomy, Space Sciences Bldg., Cornell University,
  Ithaca, NY 14853 }

\maketitle 

\begin{abstract}

The SIRTF InfraRed Spectrograph (IRS) is faced with many of the same
calibration challenges that were experienced in the ISO SWS
calibration program, owing to similar wavelength coverage and
overlapping spectral resolutions of the two instruments.  Although the
IRS is up to $\sim$300 times more sensitive and without moving parts,
imposing unique calibration challenges on their own, an overlap in
photometric sensitivities of the high-resolution modules with the SWS
grating sections allows lessons, resources, and certain techniques
from the SWS calibration programs to be exploited.  We explain where
these apply in an overview of the IRS photometric calibration
planning.

\end{abstract}

\section{THE IRS CALIBRATION CHALLENGE}

\begin{figure}[ht]
  \begin{center}
    \epsfig{file=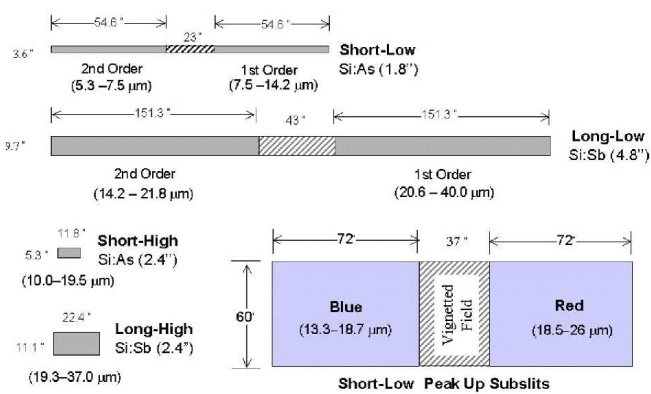, width=7cm, angle=270}
  \end{center}
\caption{Schematic representation of the IRS slits (not to scale).  The focal plane arrays are 
backside-illuminated BiB detectors in 128$\times$128 pixel format.  Detector materials, pixel sizes,
and wavelengths associated with each module are indicated.\label{slits}}
\end{figure}

The IRS is one of three science instruments in the payload of NASA's
Space Infrared Telescope Facility (SIRTF), and will provide
spectroscopy over the 5.3$-$40 $\mu$m range at spectral resolutions
$\lambda/\Delta\lambda \sim$600 between two echelle spectrographs, and
$\lambda/\Delta\lambda \sim$100 between two long-slit
spectrographs. One of the low resolution modules allows peak up
imaging at 15 and 23 $\mu$m in two subarrays, so that celestial
targets with with poorly known positions may be accurately placed in
the IRS slits.  The four instrument modules along with the cold
electronics constitute the cold assemblies located within the multiple
instrument chamber, which also houses the InfraRed Array Camera (IRAC)
and the Multiband Imaging Photometer for SIRTF (MIPS). The IRS and
MIPS share warm electronics, with savings in mass, volume, and cost.
The slit dimensions, wavelength coverage, detector materials and pixel
sizes for the four modules are schematically illustrated in
Figure~\ref{slits}.  Additional details on the performance
characteristics of the IRS may be found in Houck \& Van Cleve 1995,
Roellig et~al. 1998, and the SIRTF Observer's Manual.

\begin{figure}[ht]
  \begin{center}
    \epsfig{file=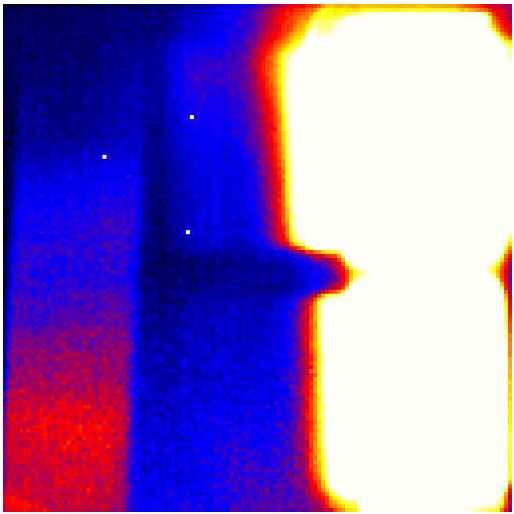, width=6.75cm}
  \end{center}
\caption{The Short Low array, showing light spilling from Peak Up subarrays on the right (red is at the top, blue
on the bottom) onto the array.  Light in the first spectral order can
be seen on the left.\label{shortlow}}
\end{figure}

One of the key design features of the IRS is the absence of moving
parts.  As a consequence, all inflight calibration activities must be
performed while light from celestial sources enters through the IRS
slits.  This impacts the measurement of dark currents and photometric
responses to both internal stimulators and external sources, as light
from the extended zodiacal background, potentially the cold ISM, and
variously from stars or other extended sources will fall on the IRS
arrays. The Short Low module will also be susceptible to the effects
of light entering through the peak up apertures and spilling from the
imaging subarrays onto the spectrally dispersed portions of the array
(Figure~\ref{shortlow}).

The challenge of shutterless operations for calibrating the IRS is
being met by a combination of:
\begin{itemize}
\item A series of intensive functional and end-to-end tests in the
laboratory, involving: the flight hardware, running from fabrication
to integration with the observatory; tests with spare arrays at
cryogenic temperatures in laboratories at Cornell University and at
the Harvard Cyclotron Laboratory; and tests of the spare pass filters
and filter assemblies in optics laboratories at Rochester University
\item data processing methods to correct for stray light
\item and the development of an iterative and flexible inflight calibration
program which stresses (i) verification and update during the In-Orbit
Checkout phase, (ii) monitoring performances and improving
calibrations by iteration over the routine mission, and (iii)
accounting for pervasive but variable background light.
\end{itemize}  

Full derivation of the spectral and Peak Up imaging flats and the
absolute flux scales will be initiated during the science verification
phase of IOC, and continued into routine operations.  The IRS has
flood illuminators in each module that can be used for measuring
detector photocurrent, linearity, noise, and stability. However, the
stimulators are non-imaging and produce a highly structured
illumination pattern, and are intended for ground-based instrument
tests without guarantees on flight performance.  Photometric
calibrations of the IRS must therefore rely chiefly on a network of
astronomical calibration sources (ACSs).  As the IRS does not have
internal sources for wavelength calibration, celestial sources will
also be observed to verify the calibrations established in the
laboratory.  The ground-based preparations and ACS selection process
are similar to those undertaken for the SWS, due to the similar
wavelength coverages and accessibility of flux densities up to several
10s of Jy by the IRS high resolution modules.

\section{IRS PHOTOMETRIC CALIBRATIONS}

\subsection{Requirements}

The IRS has a radiometric uncertainty upper limit of 5\% (van Cleve
2001).  The error budget takes into account the array dark and read
noise properties, pointing-related flux losses, and residual fringing
in the high-resolution flatfields.  As the IRS slits are narrow (see
Fig.~\ref{slits}), pointing-induced flux throughput error (assuming
the ``hard point 1'' APE of 0.4$''$) is expected to be the dominant
factor in the error budget.  This condition is familiar in SWS
calibrations.  For the high resolution modules an additional factor of
$\sim$2.5\% at 1-$\sigma$ is attributed to residual (not fully
removed) fringes in the flatfields.  Extensive testing of the
algorithm which is used to identify a source in the IRS Peak-Up field
and to subsequently place it on a slit, as well as simulations of the
stability of the telescope tracking, indicate that for most IRS
observation the 5\% radiometric accuracy requirement can be met,
following full verification of telescope pointing requirements, and
validation of pipeline processing procedures.

The radiometric accuracy requirement does not include errors arising
from uncertainties in reference data (e.g., synthetic spectral energy
distributions) for the celestial standards used in the calibrations.
Within the first year of operations, these errors are estimated to
contribute (in quadrature) an additional 10\% relative flux
uncertainty across the SL and LL spectral orders, and 5\% across the
SH and LH echelle orders.  The goal is to reduce these uncertainties
to 5\% in all orders by the end of the first year, so that the net
uncertainty across any IRS order is 7\% or less.  The relative flux
uncertainty between adjacent resolution element should be a factor of
two or more less than this by the end of the first year of operations.
The IRS does not have explicit requirements on absolute flux
calibration.  This is essentially tied to inter-order relative flux
calibration.

\subsection{Strategy}

The baseline strategy for photometrically calibrating the IRS focuses
on the spectral and Peak-Up array flats, once wavelength calibration
has been verified on-orbit.  The steps can be summarized as follows.
\begin{enumerate}
\item Coarse spatial pixel-to-pixel responsivities will be derived
for the sensitive low-resolution modules (SL, LL, and SL Peak-Up) from
dithered observations of the thermal emission from zodiacal dust
grains.  Peak-Up flats will also rely on dithered observations of a
bright diffuse reflection nebula such as NGC\,2071.
\item Refinements to the low-resolution flats, and initial derivation of the flats for the high-resolution 
modules will be achieved by smoothly scanning\footnote{Or, a
synchronized ``step-and-stare'', whereby conditioning frames are taken
during telescope motion, and exposures on the sky occur at  a fixed sky
position.} are standard stars along the slits in the cross-dispersed
direction.
\item A calibration database of early-type A dwarf stars will be built up over the mission, but a network that includes cool 
giants (G8$-$K5 III) and solar analogues is crucial due to limited
visibilities and significant uncertainties (spectral typing and
potential infrared excesses) among the A star candidates.  The Vega
phenomenon is possible around G dwarfs as well (e.g., Decin
et~al. 2000).  The spectral-typing uncertainties can be alleviated by
emphasizing stars selected for IRAC calibrations (Megeath et~al.,
these proceedings).
\item The zody-based flats can be
used to reveal significant departures in the energy distributions of
stars observed with the SL and LL modules.  The unrejected cool giants
can be used to further discriminate A and solar-type stars with
thermal infrared excesses.

\item Calibrations of the high resolution modules will be initiated solely from standard star observations, 
but will make use of stars validated with the low-resolution modules,
exploiting some overlap in the dynamic ranges of photometric responses
the modules.

\item The final flats 
will be created by median averaging the resulting 2-D images, and
using synthetic spectral energy distributions to remove the spectral
signature of the star in the dispersed direction.

\item Point source flux calibration will be established, using a calibration analysis procedure to correct point-source
 diffraction losses and higher order terms in the flatfields.  The
 ``key-wavelength'' concept used in SWS photometric calibrations will
 be employed for the IRS echelle and long-slit spectral orders.
 
\end{enumerate}

\subsubsection{The zodiacal light}

\begin{figure}[ht]
  \begin{center}
    \epsfig{file=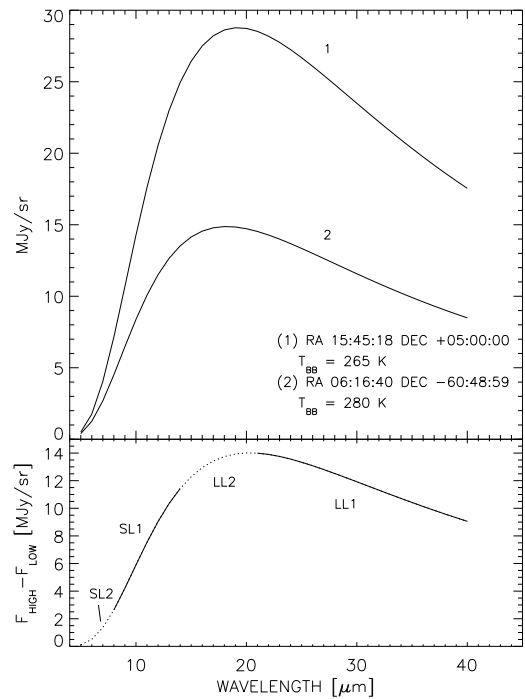, width=7cm}
  \end{center}
\caption{Predictions of zodiacal thermal emission at high and low ecliptic latitudes at the indicated positions,
computed for 15 March 2003 using a COBE/DIRBE-based model (provided by
W. Reach).  The lower curve shows the difference of high and low zody
continuum distributions, and the portions covered by the low
resolution modules (where ``SL2'' identifies Short-Low spectral order
2, etc.)\label{zody}}
\end{figure}

Approximately 9 hours of observing the zodiacal light will be
performed during IOC for flatfielding.  Dithering will mitigate the
effects of: cosmic ray events on the arrays which may not be fully
corrected in the IRS data pipeline; and spatial structure in the
zodiacal dust cloud at the arcminute scale.  The thermal peaks are
expected to differ with the solar elongation angle on any given
observation date, and will vary over time with the SIRTF orbit.  In
particular, the differences in blackbody temperatures can be
15$-$20~K, according to (geocentric) predictions of the zodiacal
thermal continuum based on COBE/DIRBE observations.\footnote{Made
available as a SIRTF proposal generation tool by W. Reach at the SIRTF
Science Center.}  Pair-differencing these observations would induce
gradients in both the SL and LL flats, most strongly in the SL module
covering 5.3$-$14.2$\mu$m on the Wien side of the energy distributions
(see Figure~\ref{zody}).  Rather than pair-differencing, therefore,
the COBE/DIRBE models must be used to divide the thermal continua out
of the observations, leaving only the instrument responses.  Averaging
the flats obtained from different pointings will alleviate temperature
uncertainties in the model.

It should be noted that uncertainties in the DIRBE-based zodiacal
models range from $\sim$5\% at the shortest wavelengths to $\sim$25\%
at the longest, but the uncertainty at 24$\mu$m has been reduced from
$\sim$20\% to $\sim$10\%, using the SWS spectrum of NGC7027 in place
of a KAO spectrum in the DIRBE calibrations.  Generally speaking,
however, confidence in the model over IRS wavelengths is lower than
desired for flatfielding purposes. Moreover, spectral effects of
potential silicate emission (Reach et~al., in preparation) are
unaccounted for in this model.  The flats resulting from IOC
observations of the zodiacal light must be regarded as coarse,
producing relative uncertainties of 20\% and 15\% across each spectral
order for the SL and LL modules, respectively.
 
\subsubsection{Stellar calibrators}

AV stars exhibit the strongest H-line spectra among all spectral types
by definition of the Harvard classification scheme.  If limited to
earliest types A0-A1, the optical-infrared line spectra are virtually
free of metal lines, which mitigates discrepancies in synthetic
spectral energy distributions imposed by limited input atomic data.
Most A dwarfs suited to IRS sensitivities have not been well observed
in the mid-infrared, and in many cases the spectral types and
luminosity classes are uncertain from lack of high quality optical
spectra.  Reliable parameters are needed for generating the models on
which the relative flux calibration will be based, eliminating those
stars with peculiar spectra (i.e., the An, Am, and Ap stars) in the
process.  Initially the IRS can benefit from the IRAC ground-based
program in which optical spectra are being obtained for the IRAC
candidate photometric calibrators to better establish their spectral
types (see Megeath et~al., these proceedings).  >From a calibration
planning perspective, {\em all} A stars will be suspect for thermal
dust emission until otherwise verified, which can involve observations
of some sources at 70$\mu$m with MIPS.  Sources with significant
excesses observable with the low-resolution modules can be
discriminated early by utilizing coarse calibrations based on the
zodiacal light measurements.  Methods for further discrimination are
described below.

Vega (A0 V) and Sirius (A1 V) are both primary MK standards which,
along with $\alpha$ Cen (G2 V), form the references for the network of
infrared calibration stars in the scheme of M. Cohen and collaborators
(see, e.g., Cohen et~al. 1992, 1996).  Both A stars should be included
among the stellar calibrators for the high-resolution modules,
allowing direct comparison with SWS observations of these stars, but
both present known and potential difficulties.

Vega has suitable photospheric fluxes, but exhibits a thermal excess
beginning near 12$\mu$m.  The excess was not confirmed with the SWS,
most likely for sensitivity reasons.  However, the excess should be
detectable with the IRS, whose LL and LH slit widths are respectively
151.3$''$ and 22.4$''$ in the spatial dimension.  Vega will be
observed but independently calibrated to investigate the strength of
its infrared dust excess in the LH slit, followed up with LL1 spectra
at short integration times, depending on the LH results. Sirius is a
spectroscopic binary, but the light of the primary completely
dominates the white dwarf companion over IRS wavelengths.  Line
velocity variations of the primary are also undetectable at IRS
spectral resolutions.  Its brightness is suitable for the LH
module,but exceeds the saturation limits for most of the SH echelle
orders.  Neither Sirius nor Vega are visible during IOC, but are in
SIRTF OPZ for at least 6 months at a time.  Visibility constrains the
number of A stars which can be observed early in the mission to only
three or four, all of which must be verified to be infrared
excess-free.

Additional calibrators will be drawn from stellar standards and lower
effective temperatures, with some overlap in makeup of the SWS
calibration program. The SWS could not take advantage of solar
analogues over most of the ISO mission due to a combination of
visibility and brightness limitations, but more of these stars should
be accessible to the IRS.  Solar analogues are emphasized in MIPS
calibrations.  Most sources will be selected from the final list of
IRAC calibrators for highest confidence in spectral typing.  The
primary calibrator designated for monitoring purposes is HR6688 (K2
III), an SWS short-wave calibrator with well-determined stellar
parameters.  Ground-based observations and Uppsala MARCS-code
synthetic spectra (Decin 2000) are available.  As a primary
calibrator, HR6688 will fill the same role as HR6705 did for SWS,
i.e. providing a reliable component in the flux calibration and a
fiducial for long-term trending on external sources.

A partial list of the primary sources is presented in
Table~\ref{starlist}.  ``Primary'' sources constitute those which have
good visibility to SIRTF, have well-determined physical properties
(confined to the usual conditions of being single, point-like,
non-variable, chromospherically quiet, etc.), and are photometrically
referencable to ground based photometry from, e.g., the ISO GBPP,
SIRTF cross-calibration IRTF, or IPAC 2MASS databases.  For these
primary sources we will seek the highest-fidelity synthetic spectra,
such as provided through collaboration between the group of
B. Gustafsson and the SWS team (see van der Bliek et~al. 1996, Decin
2000, and Decin et~al. 2001), as well as templated spectra for these
and remaining sources through SIRTF contracts with M. Cohen.

\begin{table}[!ht]
\caption{\em Partial list of primary IRS calibrators}
\leavevmode
\footnotesize
\begin{tabular}[h]{lccc}
\hline \\[-5pt]
Source & Type   &  $F_{12\mu{\rm{m}}}$ (Jy)  &  Comment \\
\hline \\
HR6606  & G9 III	& 1.70		& SL,LL,SH,LH \\
HR6688  & K2 III	& 1.40		& LL,SH,LH \\
HR6705  & K5 III	& 155.1		& LH \\
HR7018  & A0 V	& 0.19		& SL,LL,SH,LH \\
HD143187 & A0 V & 0.12 & SL, LL \\
HR7187  & G8 III	& 1.72		& SL,LL,SH,LH \\
HD46190 & A0 V & 0.120 & PU Red \& Blue \\
Neptune & Planet	& 0.05	& Red source \\
Mrk279  & AGN		& $\sim$0.2	& Red MIPS x-cal \\
\hline \\
\label{starlist}
\end{tabular}
\vspace{-2em}
\end{table}

The calibrations will not be based exclusively on an individual object
or spectral class, though the early-type A dwarfs may be increasingly
emphasized over the mission since they are more reliably templated
according to spectral type with synthetic spectra (once validated to
be free of thermal excesses) than are the yellow and red stars.

We recognize that {\em no} class of synthetic spectral energy
distributions are error-free.  A dwarf infrared photospheric line
spectra are in disagreement between theory and observations of Vega
and Sirius, for example, particularly in the lower Hu and Pf lines.
This is not yet fully understood, but generally arises from inadequate
constraints on certain contributing line formation mechanisms (e.g.,
stark broadening) in the mid-infrared (see and discussion following
presentations in these proceedings by L. Decin and M. Cohen).  At the
other extreme, tolerances must be set for potential errors in the
representation of the strong molecular bands in the cool stars.  The
SiO fundamental, spanning 7.7 to $\sim$10.5$\mu$m, is observable only
in the first spectral order of the SL module.  This and all other
bands are out of the wavelength ranges of the remaining IRS orders.
For SL1, the potential errors of the SiO band strength in models and
independent observations are gauged to be $\sim$5\% (integrated
strength), as estimated from archive-phase SWS calibrations involving
observations and the Decin (2000) MARCS model of HR6705, and ISOPHT
calibration observations of HR7341 (K1 III) and the absolutely
calibrated spectral composite from M. Cohen.\footnote{In both cases,
the SiO fundamental was represented to be too weak, by $\sim$5\%
integrated strength, leaving apparent excesses in SWS and PHT-S
spectra until corrected.}  From these lessons, cool stars should be
avoided in SL1 calibrations while the models are validated by
iteration with independently-calibrated observations.  No sources
cooler than K5 will be used directly in IRS photometric calibrations.
It should be pointed out that discrepancies in any reference data at
the 5\% level may not be resolvable by the IRS until well into
operations.

\subsection{Spectral leakage}

Red sources such as Neptune and main belt asteroids will be used to
detect flux leaking from blue wavelengths.  The stellar standards all
have negative spectral indices, and light which leaks from blue to red
wavelengths will produce apparent red excesses in science observations
if left uncorrected.  Neptune is the primary red source to search for
this effect. ISO observations and models from E. Lellouch and
M. Griffin are available.  Neptune is not visible during IOC and is
probably too bright for the LL module (chiefly LL1), however, so
asteroids as represented by thermophysical models, or Titan by SWS
spectra, may be observed with the IRS initially.  The results of
observing fast-moving solar system objects during IOC will depend on
SIRTF tracking requirements being met.

\section{DARK CURRENT MEASUREMENTS ON ORBIT}

The fact that IRS has no moving parts implies that true dark currents
cannot be measured in flight.  Dark plus low-level sky offset
measurements will be obtained several times per science campaign at
preselected positions in the north ($\alpha$=17:15:50,
$\delta$=+65:25:37, J2000) and south ($\alpha$=6:16:40,
$\delta$=-60:48:59) CVZs.  At these positions the contamination from
stars and Galactic cirrus in the mid-infrared is
minimal. Unfortunately, the areas that reach minimum cirrus emission,
and which are also targets of a number of SIRTF deep surveys, are not
in the CVZs.  The warm thermal emission from the zodiacal light will
generally dominate the sky offset, but the cooler cirrus may be
detectable at the longer wavelengths of the LL module. Consequently,
reference darks will be obtained (on less routine basis) at those
positions ($\alpha$=14:18:11.0, $\delta$=+52:29:46.4, and
$\alpha$=12:36:49.90, $\delta$=+62:12:58.0) as well. The status of the
pixel dark current will also be monitored on the regular basis,
following the behavior of the intra-order unilluminated parts of the
array.

\section{{\em{Post Mortum}} ON LESSONS LEARNED}

The preceding overview contains a number of references to aspects of
the SWS calibration program, where they most benefit photometric
calibration planning for the IRS in resources and experiences.  The
referencing is seriously incomplete, as a number of the most important
lessons learned by the SWS calibration team were learned {\em after
launch}, rather than in the pre-launch planning stages.  Several minor
and several painful modifications to the Routine Phase Calibration
Plan, distribution and focus of weekly calibration measurements, and
composition of the ACS network were made over the mission in response
to the varying sensitivities and aging of the detectors in a harsh
radiation environment (Heras et~al. 2000), to the performance of the
spacecraft (pointing and tracking), and to the learn-as-we-observe
nature of the calibration sources (leading to de-emphasis of the
asteroids, for example). Pipeline software, and capabilities of the
Interactive Analysis environment for iterative calibration analysis
and pipeline development and debug were critical in this process.  A
balance between delivery schedule and adequate down time for software
development and calibration analysis was difficult to achieve during
the cold mission, however, resulting in a schedule too heavy with
frequent pipeline and calibration update deliveries and a perpetual
PV-like workload, to the detriment of analysis depth and adequate
explanatory documentation to the science end user.  This opinion is
developed in hindsight, and an altogether different balance may be in
order for SIRTF.  The principal concern, based on the ISO experience,
must be to have the {\em tools} ready to adapt to changes in
performances after launch, as a matter of scientific survival.

\end{document}